\documentstyle[12pt,epsf]{article}
\newcommand{\eps}{\epsilon}

\newcommand{\CC}{{\cal C}}

\newcommand{\NN}{{\cal N}}

\newcommand{\LA}{\lambda}
\newcommand{\wt}{\widetilde}
\newcommand{\wh}{\widehat}

\newcommand{\wb}{\bar}
\newcommand{\de}{\hbox{deg}}

\newcommand{\be}{\begin{equation}}
\newcommand{\ee}{\end{equation}}
\newcommand{\ben}{\begin{eqnarray}\displaystyle}
\newcommand{\een}{\end{eqnarray}}
\newcommand{\refb}[1]{(\ref{#1})}

\newcommand{\sectiono}[1]{\section{#1}\setcounter{equation}{0}}

\def\J{{\bf J}}
\def\j{{\bf j}}
\def\A{{\bf A}}
\def\B{{\bf B}}
\def\C{{\bf C}}
\def\D{{\bf D}}
\def\E{{\bf E}}
\def\H{{\bf H}}
\def\X{{\bf X}}
\def\a{{\bf a}}
\def\b{{\bf b}}
\def\c{{\bf c}}
\def\x{{\bf x}}
\def\z{{\bf z}}
\def\d{{\bf \delta}}
\def\nbps{{stable non-BPS states}}

\begin{document}

{}~
\hfill\vbox{\hbox{hep-th/9907164}  
\hbox{MIT-CTP-2879}\hbox{MRI-PHY/P990722}
}\break

\vskip 2.0cm

\centerline{\large \bf Stable Non-BPS States in F- theory}  
\medskip

\vspace*{6.0ex}

\centerline{\large \rm Ashoke Sen$^{a,b}$ and Barton Zwiebach$^c$
\footnote{E-mail:  asen@thwgs.cern.ch,
sen@mri.ernet.in, zwiebach@mitlns.mit.edu}}

\vspace*{3.5ex}

\centerline{\large \it $^a$Mehta Research Institute of Mathematics}
 \centerline{\large \it and Mathematical Physics}

\centerline{\large \it  Chhatnag Road, Jhoosi,
Allahabad 211019, INDIA}

\smallskip
\centerline{\it and}
\smallskip

\centerline{\large \it $^b$International Center for Theoretical Physics}
\centerline{\large \it P.O. Box 586, Trieste, I-34100, Italy}

\medskip
\centerline{\it and} 
\medskip

\centerline{\large \it $^c$Center for Theoretical Physics}
\centerline{\large \it Laboratory for Nuclear Science,
Department of Physics}
\centerline{\large \it Massachusetts Institute of Technology}
\centerline{\large\it Cambridge, Massachusetts 02139, USA}

\vspace*{4.5ex}

\centerline {\bf Abstract}

\medskip

F-theory on K3  admits non-BPS states that are represented 
as string junctions extending between 7-branes.
We classify the non-BPS states which are 
guaranteed to be stable on account of charge
conservation and the existence of a region of moduli space where
the 7-branes supporting the junction can be isolated from 
the rest of the branes. We find three possibilities; the 7-brane
configurations carrying: (i)  the $D_1$ algebra representing 
a D7-brane
near an orientifold O7-plane, whose stable non-BPS state was
identified before, (ii) the exotic affine
$E_1$ algebra, whose stable non-BPS state seems to be genuinely
non-perturbative, and, (iii) the affine
$E_2$ algebra representing a D7-brane near a pair of O7-planes.
As a byproduct of our work we construct explicitly
all 7-brane configurations that can be isolated in a K3.
These include non-collapsible configurations of affine type.

\vfill \eject

\tableofcontents

\baselineskip=18pt

\sectiono{Introduction and Summary} \label{sx1}

Many string  
theories contain in their spectrum states which are non-BPS but are
nevertheless
stable due to the fact that they carry certain charge, and there are no
other BPS or non-BPS states of lower mass carrying the same charge into
which they can decay. A particular class of examples consist of a
configuration where a single D-$p$-brane is brought close to an
orientifold $p$-plane (O-$p$-plane). 
In this case the fundamental string
stretched between the D-brane and its image represents a non-BPS state.
Furthermore it carries charge under a U(1) gauge field living on the
D-brane, and as long as there are no other D-branes nearby, there is no
BPS state of lower mass carrying this U(1) charge into which this
non-BPS
state can decay. Thus it will represent a stable non-BPS state. Such
configurations arise in certain regions of the moduli space of the
toroidal compactification of type I string
theory.

The special case of the D7-brane $-$ O7-plane system was discussed in
\cite{9803194}. When non-perturbative effects are taken into account,
the
O7-plane splits into a pair of 7-branes \cite{9605150}. We shall use the
language of ref.\cite{9709013,9801205} to refer to the original 
D7-brane as
an
\A-type brane and to the other two 7-branes representing the 
O7-plane as \B\
and \C-type branes. In this description
the non-BPS state
can be represented as a string junction\footnote{Throughout this paper 
we shall refer to both string junctions and string networks, as
junctions.} ${\bf j}$ 
with its prongs ending on
the three 7-branes\cite{9709013,9801205}.\footnote{String junctions 
have been used in \cite{9811064} to construct non-BPS states on the
3-brane $-$ 7-brane system.} In the
limit when the
separation
between these three 7-branes is large compared to the string length
scale $l_s$, the mass of the state ${\bf j}$ can be computed
by adding up the masses of all the strings forming the
junction\cite{9803194}.

Such a configuration of 7-branes arises in the special limit
of F-theory compactification on elliptically fibered K3 \cite{9602022}
when the size of the base is large, and when the relative distances
between the three 7-branes 
representing the D7-brane O7-plane system are much smaller than
the distance between any of these three 7-branes and any of the other
twenty one 7-branes.
The stability of this
junction ${\bf j}$ can be argued as follows. 
First of all, it is
charged under a  U(1) gauge field living on the \A\B\C\  
7-brane system
and carries the minimal value of charge. Second,
there are no
BPS states on the \A\B\C\ brane system which carry  this U(1) charge. 
Indeed,
all states of the system carrying 
this U(1) charge, 
are non-BPS, 
and are 
represented by junctions $n{\bf j}$ with
$n\not=0$.
In the approximation 
where the length of each segment of the junction
is large compared to the string length scale $l_s$ so that the mass of the
junction is given
by adding the masses of the strings, the states with $n=\pm1$ are expected
to be the lowest mass ones, since in this limit  
the state $n\j$ will be represented by $n$ copies of the junction
representing the state $\j$. Thus ${\bf j}$ cannot
decay into a state living on the \A\B\C\ system.
In addition, since ${\bf j}$ 
carries a U(1) charge originating from
the three 7-brane system, it cannot decay into a state which lives
completely
on the other 21 7-branes. If it were to decay, the decay products
must include a junction ${\bf j'}$ with at least one prong extending
from the 
\A\B\C\  
system all the way to one of the 21 other 7-branes.
Since all the 21 other 7-branes are
far away, the mass of ${\bf j'}$ is necessarily much larger than
that of ${\bf j}$, and thus this decay 
is not
possible due to energetic reasons. This establishes the 
stability of ${\bf j}$.

The above argument requires the relative distance between all the 7-branes
to be large compared to the string length scale $l_s$ so that the dominant
contribution to the mass of the state 
comes from the classical mass of
the junction. However, this condition can be relaxed a little. Suppose
that the
three special 7 branes are very close to each other, so that their
separations are of the order of, or are much smaller than $l_s$. 
Then, since 
stringy corrections
could be of order $l_s^{-1}$  the computation of 
masses of the localized
non-BPS junctions living on the \A\B\C\ brane system is difficult, and 
one may not be able
to decide which of the non-BPS junctions is the 
stable one against decay
within this system.\footnote{Fortunately in the limit when the separation
between the \A\B\C\ branes are much smaller than $l_s$, 
the system can
be described by a D7-brane O7-plane system which allows us to compute the
masses, and conclude that the state carrying the minimum U(1) charge is
stable.} But there will
be at least one stable state, $-$ the
one with minimum mass carrying the U(1) charge. Since this state has mass
of order $l_s^{-1}$, 
stringy corrections will not invalidate the conclusion of
stability against
decay using the faraway branes as long as the distance
between the three isolated branes and every one of
the twenty one other 7-branes is much
larger than
$l_s$. We will therefore have a stable non-BPS
state.

In this paper we 
generalize this construction to
other limits
of F-theory compactification on K3. The basic idea will be as follows.
We first consider a limit of F-theory compactification 
where a subset of
$(24-r)$
7-branes are far away from a set of $r$ 7-branes.
We shall call the latter a set of isolated 7-branes. In 
this case
in analysing the stability of any state which lives solely on the
isolated
branes, we can forget about the existence of the other 7-branes, and
study
if the state can decay into other states living on the isolated set of
7-branes.\footnote{Again 
this argument would require the distance
between
the isolated branes and the other branes to be much larger than
$l_s$.} 
Then we study if this specific set of isolated branes contains a
non-BPS state subject to the condition that a) 
it carries a (set of)
U(1)
charge(s),
and b) 
there is no combination of BPS states living solely on the
isolated branes which
also carries the same charge quantum number(s). In this case this
isolated
brane
configuration is guaranteed to contain a stable non-BPS state carrying
charge under this U(1) 
gauge field(s). It is of course possible to find a set of
BPS
states carrying same charge quantum numbers if we include string
junction
configurations some of whose prongs end on the other 7-branes, 
but these
states are too heavy, and so it is not energetically possible for the
original non-BPS state to decay into
such states.

This construction of course does not exhaust all possible ways of
obtaining stable 
non-BPS states in F-theory compactification on K3. In
particular
one may find examples 
where there is a non-BPS state and a set of BPS
states whose total charge quantum numbers match that of this 
non-BPS
state,
but a detailed dynamical analysis
involving computation of the masses of each state shows that it is
not energetically possible for a non-BPS state to decay into the set of
BPS
states carrying the same set of chage quantum numbers. We 
do not attempt to analyze
these cases in this paper. 

Note that if we do not consider isolated 
7-brane configurations of this
kind, but consider the full set of states in F-theory on K3, then for 
every
state carrying some specific charge quantum number there is a set of BPS
states carrying the same charge quantum number, and hence every state can
in principle decay into a set of BPS states unless such decays are
forbidden due to energetic reasons. This can be seen by noting that the
full lattice of junctions for F-theory on K3 (or equivalently the Narain
lattice\cite{NARAIN} for heterotic string theory on $T^2$) can be
generated by a set of BPS states. Indeed, if one takes $E_8\times E_8$
heterotic string theory on $T^2$, then the $E_8\times E_8$ part of the
lattice is generated by the root vectors of $E_8\times E_8$ which
represent massless gauge bosons and are BPS states, whereas the four
dimensional
lattice associated with $T^2$ is generated by BPS states carrying 
unit winding
or momentum along either of the two directions.

Our analysis proceeds in several steps. In section \ref{sx2} we 
begin to
classify 7-brane configurations which can be isolated. We find two classes
of 7-brane configurations of this kind. The first class corresponds to a
configuration of 7-branes where $r$ of the 7-branes are at
finite distance of each other and the other 7-branes can be pushed all the
way to infinity.\footnote{Since in an F-theory background a constant
rescaling of the metric does not destroy the solution, we can start from
this solution and go to other configuraions where the distance between the
$r$ 7-branes are large or small compared to $l_s$.} In this case we have a
non-singular background describing only these $r$ 7-branes; and we call
these {\it properly isolated 7-brane configurations}. 
The other class consists
of 7-brane configurations where $r$ of the 7-branes are within a finite
distance of each other, and the other 7-branes are at a distance larger 
than $L$ for some large number $L$. In this case, however, we cannot take
the $L\to \infty$ limit and push the other branes all the way to infinity,
because in this limit the string coupling constant vanishes everywhere in
the region within finite distance of the isolated branes. We call these
{\it asymptotically isolated 7-branes}. 
We find that the monodromy
around a properly isolated brane
configuration must be an elliptic or a parabolic element of the SL(2,Z)
S-duality group, whereas the monodromy around an asymptotically isolated
7-brane must be a parabolic element. In particular, we see that 7-brane
configurations with hyperbolic monodromies cannot be isolated.

In section \ref{sx3} we look for explicit examples of isolated 7-brane
configurations using the list of 7-brane configurations found in
ref.\cite{9812209}. In this list we find that the 7-brane configurations
$\E_6$, $\E_7$, $\E_8$, $\H_0$, $\H_1$, $\H_2$, $\wh \E_n$ ($1\le n\le
9$), $\wh{\wt \E}_0$, $\wh{\wt \E}_1$, and $\D_n$ ($0\le n\le 4$) satisfy
the necessary conditions for being properly isolated 7-brane
configurations, and we show that these conditions are also sufficient by
explicitly constructing the properly isolated 
7-brane configurations of
these
types. The analysis is more complicated for asymptotically isolated
7-brane configurations, and we have not attempted a thorough study of all
the configurations listed in ref.\cite{9812209} to see which of them can
be asymptotically isolated. However, we show the  
existence of 7-brane
configurations of this kind based on $\A_n$ ($n\ge 0$) and $\D_n$ ($n\ge
5$) type configurations.

In section \ref{sx4} we examine each of the examples of isolated 7-brane
configurations found in section \ref{sx3} and look for stable non-BPS
states in this system. The basic idea has already been explained before: we
demand that we have one or more non-BPS junctions carrying a 
(set of) U(1)
charge(s), and that there is no combination of BPS states living on the
isolated brane system carrying the same (set of) U(1) charge(s).
According to our previous argument, this
would guarantee that there is at least one stable non-BPS state on the
isolated 7-brane system as long as the faraway branes are much farther  
than $l_s$ away. We find that there are three 7-brane configurations 
satisfying
these constraints $-$ $\D_1$, $\wh \E_2$ and $\wh{\wt \E}_1$. Of these
$\D_1$
represents a D7-brane near an O7-plane, and $\wh{\E}_2$ 
represents a
D7-brane near a pair of O7-planes. The existence of possible non-BPS
states in these configurations could be argued by working in the
orientifold limit. On the other hand, $\wh{\wt \E}_1$ seems to represent a
genuinely new example 
as it gives a non-BPS state with no simple perturbative
interpretation. 

In section \ref{sx5} we study the possibility of obtaining non-BPS
states on non-isolable 7-brane
configurations. We construct several examples of 7-brane configurations
containing string junctions which are stable
against decay within the given 7-brane configurations. But since these
7-brane configurations are not isolable, there are other 7-branes nearby,
and these junctions could be
unstable against decay into string junctions which have one or more prongs
ending on 7-branes outside this system.

As we were preparing to submit this paper, an interesting work 
by Y.~Yamada and
S.~K. Yang appeared \cite{9907134} which also gives an explicit 
construction of
the affine exceptional brane configurations. This substantially 
overlaps with section
\ref{sx32}.

\sectiono{Constraints on Isolated Configurations} \label{sx2}

Various configurations of $(p,q)$ 7-branes were studied in
refs.~\cite{9812028,9812209}. In this section we shall analyse the
conditions under which a given set of 7-branes can be
considered in isolation.
This issue arises because we
want to consider subsets of the
configuration of 24 7-branes
describing F-theory on K3 \cite{9602022}.
We try to take an
appropriate limit in the parameter space where a chosen set of
7-branes is far away from all the other 7-branes.
We shall say
that the chosen set of 7-branes can be isolated
if it is possible to
consider the limit in which
the largest distance between any two members of the chosen set can be
made
small compared to the
distance between the chosen set and
any of the other 7-branes.
As discussed in the introduction,
only for
brane configurations that can be considered
in isolation we can reliably ascertain the existence of stable
non-BPS states.

Our analysis of stability of non-BPS states 
is based
on classical considerations, and thus our results are valid in the 
limit  when the size of the $S^2$ base transverse to
the seven branes
is taken to
be large compared to the string length $l_s$.
In particular, with the help of an overall rescaling of the metric, we
choose the size of the base to be sufficiently large so that the
distance between the isolated branes and any of the other branes is
large compared to the string length $l_s$. In this limit, as explained
in the introduction, a stable non-BPS state on the isolated 7-brane
configuration
cannot be rendered unstable by the presence of the faraway branes.

Our analysis will consist of two steps. In the first step we
find constraints on monodromies that can appear around an
isolated configuration by requiring that
all other 7-branes are at large coordinate distance away from the
isolated branes. In particular, we find that
hyperbolic
monodromies are not allowed. Since large coordinate
distance does not always correspond to large distance measured in the
relevant metric, in the second step
we impose the
condition that when distances are measured in the appropriate metric, the
isolated configuration is still far away
from the remaining branes.

\subsection{Constraints on monodromies for isolated configurations}
\label{sx21}

A configuration of 7-branes in F-theory is described by specifying a
pair of polynomials $f$ and $g$ in $z$, $-$ the complex coordinate
parametrizing the
space transverse to the 7-brane. $f$ is a polynomial of degree $8$
and $g$ is a polynomial of degree $12$. We define:
\be \label{exone}
\Delta = 4 f^3 + 27 g^2\, .
\ee
Then the dependence of the axion-dilaton modulus
$\tau(z)$ on the
transverse coordinate $z$ is given by:
\be \label{extwo}
j(\tau(z)) = {4 \cdot (24 f)^3\over \Delta}\, .
\ee
$j(\tau)$ blows up at the zeroes of $\Delta$. These are the locations of
the 7-branes.

We can now make more precise
what we
mean by
isolating a set of branes. For this note that $f$ and $g$ are
labelled by a set of $22$ parameters $\xi_i$. We shall consider
the cases where we can focus on a one dimensional subspace
$\xi_i(\lambda)$, parametrized by $\lambda$.
We say that we can isolate a set of branes if as we
take the limit $\LA\to 0$ the parameters $\xi_i(\lambda)$
of $f$ and $g$ flow in such a way that
a set of $r$ roots of $\Delta$ remain at finite points in the $z$
plane,  while the others move off to 
infinity. In that case for a finite
but sufficiently small $\LA$, by using  scaling and translation in
$z$, we can ensure that $r$ of the zeroes of $\Delta$,
$-$ which we shall associate with the location of the isolated branes,
$-$
are within the unit disk centered at the origin, and
the faraway branes are outside a circle of radius $L$, also
centered at the origin, where $L$ is some arbitrary but fixed
large number.

The above definition will describe an isolated set of $r$ 7-branes in
the sense described earlier if we can show that finite (infinite)
coordinate distance in
the $z$-plane corresponds to finite (infinite) distance
measured in
the  metric used for computing the 
mass of a $(p,q)$ string
stretched along a geodesic\cite{GREENE,9608005,9608186}. 
This must be true for all possible values of $(p,q)$. 
This constraint will give
additional restrictions on the form of $f$ and $g$. 
We shall derive these
constraints in the next subsection. 

\medskip
Let $z_i$ denote the
positions of the isolated branes
and $\wt z_i$ the
position of the { faraway} branes.
By using the {additional}
freedom of simultaneously rescaling
$f$ and $g$ by constants
$\gamma^2$ and $\gamma^3$ respectively, we can bring $\Delta$ to the
form:
\begin{equation}
\label{edelta}
\Delta = \prod_{i=1}^r (z-z_i) \prod_{i=1}^{24-r} (1 - {z\over
\wt z_i})
\end{equation}
{As mentioned before, for sufficiently small $\lambda$
we have $|z_i| < 1$ for all $i=1, \cdots r$ and
$|\wt z_i|> L$ for all $i=1, \cdots , 24-r$.  }
As we travel (in the {\it clockwise direction}) 
around the set of
isolated
branes, the modulus $\tau$ undergoes an $SL(2,Z)$ transformation:
\be \label{exfour}
\tau \to {a\tau + b\over c\tau + d}\, .
\ee
We shall call the matrix $\pmatrix{a & b\cr c & d}$ the 
monodromy matrix
$K$ around the isolated set of 7-branes.
Our first task will be to
compute the possible monodromy matrices $K$ around the isolated
branes.\footnote{Throughout this and the next section 
we shall determine the
monodromy matrix only up to an SL(2,Z) conjugation, unless mentioned
otherwise.}
For computing $K$  we can use any contour surrounding the isolated
branes
but not enclosing any faraway brane.

We now examine the behavior of $f(z)$ and $g(z)$.
We shall call a parameter in $f$ or $g$ {\it small}, {\it finite} or
{\it large}, if
taking
$\LA$ to zero requires the parameter to go to zero,
remain bounded within a circle of finite radius in the 
complex plane, or
grow indefinitely, respectively.
We write $f$ and $g$ in the form
\be \label{exff}
f(z)=F
\prod_{i=1}^{d_f} (z-u_i) \prod_{i=1}^{8-d_f} (1 -{z\over \wt u_i})\, ,
\ee
\be \label{exgg}
g(z)=G \prod_{i=1}^{d_g} (z-v_i) \prod_{i=1}^{12-d_g} (1 -{z\over \wt
v_i})\, ,
\ee
where we have introduced parameters $u_i, \wt u_i, v_i , \wt v_i$,
entering
as zeroes of $f$ and $g$, and parameters $F,G$ entering as overall
coefficients.
The parameters
$u_i$ ($1\le i\le d_f$)  are assumed to be either
{\it small} or {\it finite},
while the
$\wt u_i$ ($1\le i\le (8-d_f)$) are assumed to be {\it large}.
Similarly, $v_i$ ($1\le i\le d_g$) are small
or finite, and  $\wt v_i$ ($1\le i\le
(12-d_g)$) are large.
The above expressions
for $f$ and $g$ are completely general.
Our analysis will require investigating the nature of
the parameters $F$ and $G$.
Note that having used scaling to fix the
distribution of branes implicit in $\Delta$ and the overall 
normalization of $\Delta$, we no
longer
have any further scaling freedom to set $F$ and/or $G$ 
to specific values. The parameters $F$ and $G$ can be small, finite or
large. 
The numbers $d_f$ and $d_g$ indicate the degrees of $f$ and $g$
respectively, when
we use only the factors associated with 
the small and finite roots.

{We shall now introduce two contours
that will help in the analysis.
Since we have a bounded number of
small and finite roots one can define a finite length $R/2$ which is
the magnitude of the largest finite root (of $f$, $g$ or $\Delta$) in
the
limit
$\LA\to 0$. It then follows that, for sufficiently small $\LA$,
the circle ${\cal C}$ of finite radius $R$
contains all small and finite roots. Moreover,
this circle is at a finite distance from
all the small and
finite roots of $f$, $g$ and from all the isolated
branes.
In addition, if $\wt L$ denotes the magnitude 
of the smallest large root of $f$, $g$ or $\Delta$, then
for sufficiently small $\LA$ we can choose another circle ${\cal C}'$,
of radius $R'$ such that $R'/R$ {\it and} $\wt L/R'$ 
are arbitrarily
large. This circle is both far outside the finite and small roots, and
far
inside the large roots and the faraway branes. As mentioned before we
can
calculate the monodromy $K$ using
${\cal C}$ or ${\cal C}'$, since each of them only 
encloses the isolated
branes (note that
crossing zeroes of $f$ and $g$  does not affect
the monodromy; however in deforming $\CC$ to $\CC'$ we do not cross any
zero of $f$ or $g$).}

By construction $\Delta$
takes finite values on ${\cal C}$
(see (\ref{edelta})).
In addition, on ${\cal C}$, we have $f\sim F$ and $g\sim G$, by which
we mean that $f /F$ and $g/G$ are both finite and nonzero.
Since
$\Delta=4f^3+27 g^2$, we see that
both
$F$ and $G$ cannot be small parameters.
If $F$ is
large $G$ also must be large and vice versa. Thus we need to consider
the
following cases separately: 1) $F$ small, $G$ finite 2)
$F$ finite, $G$ small 3) $F$
and $G$ both finite, and
4) $F$ and $G$ both large.

\begin{enumerate} \item $F$ small, $G$ finite: In this case on the
contour
${\cal C}$, $\Delta\simeq 27 g^2$, and hence from
\refb{exone}, \refb{extwo} we see that $j(\tau)\simeq 0$. This gives
$\tau \simeq e^{2i\pi/3}$
up to an SL(2,Z) conjugation.
The monodromy $K$
around ${\cal C}$ must leave this value of $\tau$ fixed.
This gives
$K= \pm(ST)^{\pm 1}$ or $K =\pm 1$, where we define:
\be \label{exthree}
S = \pmatrix{ 0 & -1\cr 1 & 0}, \qquad T 
= \pmatrix{ 1 & 1\cr 0 & 1}\, .
\ee

\item $F$ finite, $G$ small: Now $\Delta\simeq 4 f^3$ along ${\cal C}$.
This
gives $j(\tau)\simeq (24)^3$, and
$\tau\simeq i$ along ${\cal C}$. The monodromy around ${\cal C}$ which
must
leave
this
value of $\tau$ fixed is either $K= \pm S$ or $K = \pm 1$.

\item $F$ and $G$ finite: In this case we shall  compute $K$
using the contour
${\cal C}'$. On ${\cal C}'$ the functions $f$ and $g$ are
well approximated by their overall coefficients $F$ and $G$
together with
the factors containing the small and finite roots. There are
several subcases to be considered:

\begin{enumerate}
\item $3d_f\ne 2d_g$:  Depending on whether $3d_f>2d_g$ or $3d_f<2d_g$,
$f$ or $g$
will be
the dominant contribution to $\Delta$. In the first case $\tau\simeq i$
on $\CC'$, and
$K = \pm S$ or $K= \pm 1$. In the second case $\tau\simeq e^{2\pi i/3}$
on
$\CC'$ and  $K = \pm (ST)^{\pm
1}$ or
$K = \pm 1$.

\item $3d_f=2d_g$. In this case there are two possibilities: $r<3d_f$
and
$r=3d_f$.  Since $\Delta=4f^3+27 g^2$, and
$f$ and $g$ are approximated by polynomials of degree $d_f$ and $d_g$
respectively, the approximation to $\Delta$ is a polynomial of degree
at most $3d_f(=2d_q)$. Hence $r>3d_f$ is not possible.

\begin{enumerate}

\item $r<3d_f$: In
this case on $\CC'$
we have
\be \label{exfive}
j(\tau) = 4(24f)^3/\Delta \sim z^{3d_f-r}\, . 
\ee
Since $j(\tau)$ is large on $\CC'$, we can use
\be \label{exsix}
j(\tau) \sim e^{-2 \pi i \tau}\, ,
\ee
to conclude that
\be \label{exseven}
\tau \simeq - (3d_f -r) {1\over 2\pi i} \ln z + \hbox{constant}\, .
\ee
This implies
$K= \pm
T^{3d_f-r}$. 
Note that the power of $T$ is positive. 

\item  $r=3d_f$: In this case
$f^3/\Delta$ goes 
to some constant on $\CC'$. Thus $\tau$ goes to a constant. Since
$f^3/\Delta$ and $g^2/\Delta$ are both 
non-vanishing and finite on $\CC'$, the constant $\tau$
is
not (an SL(2,Z) conjugate
of) $i$, $e^{2\pi i/3}$ or $i\infty$. The monodromy along ${\cal C}'$
must leave fixed this constant value of $\tau$.
The only possibility is $K= \pm
1$. This case is identical
to (several copies of) the $D_4$ case \cite{9605150}.

\end{enumerate}
\end{enumerate}

\item $F$ and $G$
large. In this case along  ${\cal C}$ the function $f$ is large but
$\Delta$ is finite.
Thus we can use eq.\refb{exsix} to conclude that Im($\tau)$
must be large along $\CC$.   The
monodromy around ${\cal C}$
must preserve this condition. This gives  $K=\pm T^{k}$ for some 
integer $k$ ($-\infty<k<\infty$). 

\end{enumerate}

Note that in cases $1,2$ and $3$,
the functions $f$, $g$ and
$\Delta$ remain well defined in the $\lambda\to 0$ limit,
since they all
approach finite values for finite $z$
in the $\lambda\to 0$ limit, and each of them only has a
finite number of isolated zeroes in the finite $z$ plane. In these
cases even when we set
$\lambda=0$,
$\tau(z)$ is well defined in the sense that it is finite (with
$Im(\tau)>0$) for finite
$z$ except at isolated points which are the locations of the 7-branes.
On the other hand,
in case 4, $f$ and $g$ do not approach a
finite value for finite $z$ as $\lambda\to 0$, since
the multiplicative factors
$F$ and $G$ blow up in this
limit. Since $j(\tau) \sim f^3/\Delta$ and $\Delta$ is bounded
for any finite $z$,  for
$\lambda=0$ we will have
$Im(\tau)=\infty$ at every
finite point in the $z$ plane except at  the zeroes of $f$.
Since the
string coupling constant is given by the inverse of the imaginary
part of $\tau$, we see that in this case the string coupling approaches
zero at
every finite point in the
$z$ plane except at the zeroes of $f$.
This is a singular configuration. 
But since in order to get 7-brane configurations admitting stable
non-BPS
states we do not need to actually set $\lambda=0$, but only 
need to take
$\lambda$ sufficiently small, even these kind of 7-brane configurations
are potentially good candidates for admitting stable non-BPS states in
their spectrum.

\subsection{Constraints from large distance separation} \label{sx22}

Let us now study the constraints
coming from the requirement that
finite
(large) separation in the coordinate $z$ correspond to finite (large)
separation measured in the appropriate metric used in 
computing the mass
of a string junction 
living on the 7-brane system.
Up to an
arbitrary constant multiplicative factor, this metric used in computing
the mass of a $(p,q)$ string is given by \cite{GREENE}: 
\be \label{emetric}
ds_{p,q} =|p + q\tau| |\eta(\tau)|^2 \prod_{i=1}^r |z - z_i|^{-{1\over 12}}
\,|
dz|\, ,
\ee
where $z_i$, as before, denote the locations of the isolated 
7-branes. First
let us consider the cases 1-3 discussed above. In these cases we 
can set $\lambda=0$ from the beginning. Now 
$\tau$
is finite everywhere in the $z$-plane except at the locations of the
7-branes, and possibly at $z=\infty$. The metric of a $(p,q)$-string is
known to be finite near a $(p,q)$-seven brane. Thus it is finite at all
finite points in the $z$-plane, and hence finite coordinate distance
will
correspond to finite distance measured in the metric
\refb{emetric}.\footnote{The metric of a $(p,q)$-string may have mild
divergence near a $(p',q')$ 7-brane for $(p,q)\ne (p',q')$, 
but we can
always choose contours
which avoid such points.} Thus if we can show
that infinite coordinate distance corresponds to infinite distance when
measured in the metric \refb{emetric}, 
for any $(p,q)$ values, 
we would have
shown that finite (large) separation
in the
$z$-coordinate system corresponds to finite (large) separation in the
metric
\refb{emetric}. The main issue here is whether the point $z=\infty$ is
infinite or finite distance away from finite points in the $z$-plane.
For
this we note that for large $|z|$, \refb{emetric} reduces to
\be \label{exm1}
ds_{p,q} \simeq |p + q\tau| |\eta(\tau)|^2\, |z|^{-{r\over 12}} |dz| \, .
\ee
In cases 1, 2, 3(a) and 3(b)(ii), $\tau$ approaches a finite value as
$|z|\to\infty$.
{}From eq.\refb{exm1} we see that in this case the point $z=\infty$ is
infinite distance away for\footnote{Similar results were obtained in
ref.\cite{GREENE}.} 
\be \label{exyz1}
r\le 12\, .
\ee
In case 3(b)(i) the monodromy is $\pm T^k$ with $k\equiv (3d_f-r)>0$.
In this case $\tau\to i\infty$
as $|z|\to\infty$. More precisely,
\be \label{exm2}
j(\tau)\sim e^{-2\pi i\tau} \sim z^k\, .
\ee
This gives, for large $|z|$,
\be \label{exm3}
\eta(\tau) \sim e^{2\pi i \tau/24} \sim z^{-{k\over 24}}\, .
\ee
Thus \refb{exm1} takes the form:
\be \label{exm4}
ds_{p,q} =|p+q\tau| |z|^{-{r+k\over 12}} |dz| \, .
\ee
Since $|p+q\tau|$ either approaches a finite value or 
grows logarithmically as 
$|z|\to\infty$, we see that the point $z=\infty$ is at 
infinite distance
measured in the metric \refb{exm4} if
\be \label{exm5}
r+k\equiv 3d_f \le 12\,.
\ee

Finally we turn to case 4. In this case Im($\tau$) is large for finite
$z$, and hence
\be \label{exx1}
j(\tau) \sim e^{-2\pi i\tau} \sim F^3
\prod_{i=1}^{d_f}(z-u_i)^3 
\prod_{i=1}^r(z-z_i)^{-1}\, .
\ee
This gives
\be \label{exx2}
\eta(\tau) \sim e^{2\pi i\tau/24} \sim F^{-{1\over 8}}
\prod_{i=1}^{d_f}(z-u_i)^{-{1\over 8}} 
\prod_{i=1}^r(z-z_i)^{{1\over 24}}
\, .
\ee
Hence
\be \label{exx3}
ds_{p,q} \sim |p+q\tau| |F|^{-{1\over 4}}
\prod_{i=1}^{d_f}|z-u_i|^{-{1\over 4}} |dz|\, .
\ee
Since $\tau\simeq -{3\over 2\pi i}\ln F$ is almost constant, 
we see from
eq.\refb{exx3} that the distance between two finite points in the $z$
plane is small compared to the distance between a finite point and a
point
at large $|z|<<\wt L$ provided 
\be \label{exx4}
d_f\le 4\, .
\ee

In the next section we shall look for 7-brane configurations
satisfying
all the constraints found in this section. For this analysis it will be
useful to divide the possible set of isolable 7-brane configurations
into
two classes. Since in cases 1, 2 and 3, we can actually set $\lambda=0$
and get a well defined function $\tau(z)$, we shall call these
configurations
properly isolated 7-brane configurations. 
On the other hand in case 4 we
only get isolated 7-brane configurations for small but non-zero
$\lambda$.
If we try to push all the other 7-branes all the way to infinity by
taking the $\lambda\to 0$ limit,
$Im(\tau)$ blows up for all finite $z$ except at isolated points. We
shall
refer to these configurations as asymptotically isolated 7-brane
configurations.
As discussed earlier, both kinds of isolated 7-brane
configurations are potentially relevant for finding stable non-BPS
states.

\section{Constructing the Isolated Configurations} \label{sx3}

Our analysis so far
gives constraints on
monodromies for isolated 7-brane
configurations, but does not guarantee that given a 7-brane
configuration with one of these monodromies, it can always be isolated.
We can now use Table 5
of \cite{9812209} to identify the brane configurations
giving such monodromies.
Brane configurations were classified
by their monodromy matrices $K$ into elliptic $(|Tr(K)|<2)$, parabolic
$(|Tr(K)|=2)$ and hyperbolic ($|Tr(K)|>2$) type. $\pm S$ and $\pm
(ST)^{\pm1}$ are examples of elliptic monodromy, whereas $\pm T^k$ for
any integer $k$ are examples of parabolic monodromy. From our analysis
we
see
that only 7-brane configurations with elliptic and 
parabolic monodromies
can be possibly isolated.

Let us first consider 7-brane configurations which can be properly
isolated.  In this case we can
have elliptic
monodromies coming from cases 1, 2 or 3(a), and parabolic monodromies
coming from cases 1, 2, 3(a) or 3(b) of section \ref{sx21}. 
In the
elliptic cases equation
\refb{exyz1} must be satisfied.
{}From table 5 of ref.\cite{9812209} we see
that with fewer than 12 7-branes we have the configurations:
$\E_6$, $\E_7$, $\E_8$, $\H_0$, $\H_1$, and $\H_2$.
Parabolic cases coming from cases 
1, 2, 3(a) and 3(b)
have monodromy $\pm T^k$
with $k\ge 0$. Furthermore, we need to satisfy
\refb{exm5}.
In this class we find
$\wh \E_N$
($9\ge N\ge 1$),
$\wh{\wt
\E}_1$, $\wh{\wt \E}_0$, 
 and $\D_N$ ($4\ge N\ge 0)$. We shall
explicitly construct $f$ and $g$ for each of these elliptic and
parabolic configurations, thereby proving
that all these configurations can be properly isolated. Of these, 
the ones with elliptic monodromy, and the $\D_4$ configuration 
can be collapsed to a
single point, but none of the other configurations with parabolic
monodromy can be collapsed to a single point.

Asymptotically isolated configurations originating 
from case 4 in section \ref{sx21} 
can have
monodromy $\pm T^k$ where $k$ is any integer. There are several
configurations with monodromies of this form. For example, after
excluding
the properly isolated configurations, we have $\A_n$ ($n\ge 0$), $\D_n$
($n>4$), $\wh \E_n$ ($n>9$), three copies of $\D_0$ etc.\footnote{Two
copies of $\D_0$ is equivalent to $\wh \E_1$\cite{9812209}.} But in this
case a complete analysis of which of these configurations can actually
be
isolated is more difficult, since these configurations can be reached
only
as a limit. However, we do give proof of existence of a class of such
configurations based on resolutions of $\D_N$ ($N\ge 5$) 
and $\A_N$ ($N\ge
0$) singularities.

We begin our analysis with the properly isolated configurations.

\subsection{Elliptic cases} \label{sx31}

We now consider the configurations with elliptic monodromies
having less than twelve seven-branes. These all correspond
to Kodaira singularities and we will see that they can be 
isolated.  The constraints
given below on the polynomials $f$ and $g$ were listed in
\cite{9903215}, which also extended the work of \cite{9608047}
on the construction of curves for ${\cal N} =2$ 
supersymmetric
four dimensional gauge theories with global exceptional symmetries.

$\E_6$: This contains
eight 7-branes and has monodromy
$-(ST)^{-1}$.\footnote{Again in this section we continue to
use monodromies that are only fixed up to SL(2,Z).}
The fact
that there are eight 7-branes means that for a 
properly isolated 7-brane
configuration of this type, $\Delta$ is a
polynomial of degree 8. The monodromy matrix leaves fixed the point
$\tau=e^{2i\pi/3}$.
Hence far away from the seven brane configuration $\tau$ must
approach $e^{2i\pi/3}$,
and $j(\tau)$ must vanish.
This shows that $f^3$ must be a polynomial of degree $<8$. This gives
the
following constraints:
\be \label{exnine}
\de(f)\le 2, \qquad \de(g)=4\, .
\ee
There are no further constraints on $f$ and $g$.
In order to prove that this really describes an $\E_6$ configuration we
note that since the parameter space is connected, it is enough to show
that any one point in the parameter space describes an $\E_6$
configuration. If we consider the special case where $f=0$ and $g=z^4$,
then this describes an $\E_6$ singularity, and hence 
certainly represents
an $\E_6$ type 7-brane configuration. Thus any pair
of $f$ and $g$ satisfying \refb{exnine} gives a properly isolated
7-brane
configuration of $\E_6$ type. This 
shows that the $\E_6$ configuration can
be isolated.

$\E_7$: This contains nine 7-branes, 
and has monodromy 
$S$.  
Thus $\Delta$ is a polynomial of degree 9, and $\tau$ 
approaches $i$ far away
from the
7-branes. The latter condition tells us that $g^2/\Delta$ must vanish
sufficiently far away from the 7-branes, and hence $g^2$ must have
degree
$<9$. This gives the following necessary conditions for a
properly isolated 7-brane configuration of $\E_7$ type:
\be \label{exten}
\de(f)=3, \qquad \de(g)\le 4\, .
\ee
Taking $f(z)=z^3$ and $g(z)=0$ we get a collapsed configuration with
$\E_7$
singularity. This shows that eqs.\refb{exten} are also sufficient for
getting an $\E_7$ configuration.

$\E_8$: This has ten 7-branes and has monodromy $(ST)$. Following the
same
analysis as the $\E_6$ case we find that the necessary and sufficient
condition for having a properly isolated 7-brane configuration of this
type is:
\be \label{ex11}
\de(f)\le 3, \qquad \de(g)=5\, .
\ee

$\H_0$: This has two 7-branes and has monodromy $(ST)^{-1}$. Following
the
same
analysis as the $\E_6$ case we find,
\be \label{ex12}
\de(f)\le 0, \qquad \de(g)=1\, .
\ee
Here $\de(f)\le 0$ means that $f$ can be either a constant or zero.

$\H_1$: This has three 7-branes and has monodromy $(S)^{-1}$. Following
the
same
analysis as the $\E_7$ case we find,
\be \label{ex13}
\de(f)=1, \qquad \de(g)\le 1\, .
\ee

$\H_2$: This has four 7-branes and has monodromy $-(ST)$. Following
the
same
analysis as the $\E_6$ case we find,
\be \label{ex14}
\de(f)\le 1, \qquad \de(g)=2\, .
\ee
This finishes all the elliptic cases. We now turn to the parabolic
cases.

\subsection{Properly isolated parabolic cases} \label{sx32}

Here we must deal with two series of configurations.
One series carries affine exceptional algebras and the other
series carries orthogonal algebras. We begin with:

\subsubsection{The exceptional series $\wh\E_n$}

$\wh{\E}_9$: This has twelve 7-branes and its monodromy is the identity.
This corresponds to
\be \label{ex15}
\de(f)\le 4, \qquad \de(g) \le 6\,,  \qquad \de(\Delta) = 12.
\ee
Indeed, choosing arbitrary fourth order and sixth order polynomials
for $f$ and $g$ respectively, it is clear that for large $z$ we can get
arbitrary constant values of $\tau$. Only $K = \pm 1$ can leave
such values invariant. On the other hand, it follows from the
arguments of ref.\cite{9812209}, section 4 that twelve 7-branes cannot
produce
 $K= -1$. We must therefore have $K=1$.
This configuration arises from two copies of the
$\D_4$ case discussed in \cite{9605150}.

If we define the coefficients $f_k$, $g_k$ and $d_k$ through
the relations:
\ben \label{ex17}
\overline f (z) & \equiv &
 - (4)^{1/3} f(z) = \sum_{k=0}^4 f_k z^k \nonumber \\
\overline g (z) & \equiv & (27)^{1/2} g(z) = \sum_{k=0}^6 g_k z^k
\nonumber \\
\Delta & \equiv & -\overline f^{\,3}+\overline g^{\,2} =\sum_{k=0}^{12}
d_k z^k\, ,
\een
then, in order to get $\wh{\E}_9$ one must have  $d_{12}\not= 0$,
which requires
$g_6^2-f_4^3$ to be non-zero.

\bigskip
\noindent
$\wh \E_n$ ($2\le n\le 8$), $\wh{\wt \E}_1$, $\wh{\wt \E}_0$: 
The $\wh \E_n$
configuration has $(n+3)$ seven branes,
and has monodromy $T^{9-n}$. Thus for a properly isolated brane
configuration of this kind, $\Delta$ is a polynomial of degree $(n+3)$,
and $j(\tau)$ for large but finite $z$ behaves as
\be \label{ex16}
j(\tau) \sim z^{9-n}\, .
\ee
Using \refb{exone}, \refb{extwo} we see that $f^3\sim z^{12}$, and
$g^2\sim z^{12}$ for large $|z|$. 
Thus $f$ is a polynomial of degree 4
and $g$ is a
polynomial of degree $6$. Let us 
introduce the coefficients of expansion
$f_k$, $g_k$ through the relations \refb{ex17}.
Using the freedom of shifting $z$, and the freedom
of scaling $f$ and $g$ by $\gamma^2$ 
and $\gamma^3$ respectively for any
complex
number $\gamma$, we set
\be \label{eextwo}
f_0=0, \qquad
f_4=1\, .
\ee
This still leaves a residual rescaling freedom where we scale
$g$ by $-1$ and leave $f$ unchanged; we shall use this later. We are
also
left with the freedom of scaling $z$ by a complex number $K$ together
with
a compensating scaling $f\to K^{-4} f$, $g\to K^{-6} g$ so as to
preserve
the $f_4=1$ condition. We shall also make use of this later. 

In order to get the $\wh \E_n$ configuration, we need to ensure that the
coefficients $d_k$ defined in eq.\refb{ex17} vanish for $k\ge (n+4)$,
and
that $d_{n+3}$ does not
vanish. We shall begin by describing the solution for $\wh \E_2$. First
of
all, requiring the coefficient $d_{12}$ 
to vanish we get $g_6=\pm 1$. We
can now use the residual scaling freedom $g\to -g$, $f\to f$ to set,
\be \label{ex18}
g_6 =1 \equiv \wh g_6\, .
\ee
Now by equating the coefficients of $d_{11},\ldots d_6$ to 0,
we get\footnote{This calculation
is straightforward but
complicated, and has been done with the help of the algebraic
manipulator programme MAPLE.}:  
\ben \label{ex19}
d_{11}=0 : \qquad g_5 &=& {3\over 2} f_3 \equiv \wh g_5\, ,
\een
\ben \label{ex20}
d_{10}=0 : \qquad g_4 &=& {1\over 2} (3 f_2 + 3 f_3^2 - g_5^2) \nonumber
\\
&=& {3\over 2} f_2
+{ 3\over 8} f_3^2 \equiv \wh g_4\, .
\een
\ben \label{ex21}
d_9=0 : \qquad g_3 & = & {1\over 2} (3f_1+6f_3f_2+f_3^3-2g_4g_5) 
\nonumber
\\
&=& {3\over 2} f_1+ {3\over 4} f_2f_3
-{1\over 16} f_3^3 \equiv \wh g_3\, .
\een
\ben \label{ex22}
d_8=0 : \qquad g_2 & = & {1\over 2} (6 f_1 f_3 + 3f_2^2 + 3 f_3^2 f_2 -
2g_3 g_5 -g_4^2)\nonumber \\
&=& {3\over 8} f_2^2 - {3\over 16} 
f_2 f_3^2 + {3\over 4} f_1 f_3 + {3\over 128} f_3^4 \equiv \wh g_2
\een
\ben \label{ex23}
d_7=0 : \qquad g_1 & = & {1\over 2} (6 f_1 f_2 + 3 f_1 f_3^2 + 3 f_3
f_2^2
- 2g_2 g_5 -
2g_3 g_4) \nonumber \\
&=& {3\over 4} f_1 f_2 - {3\over 16}
f_2^2 f_3 + {3\over 32} f_2 f_3^3 - {3\over 16} f_1 f_3^2 - {3\over 256}
f_3^5 \equiv \wh g_1\, .
\een
\ben \label{ex24}
d_6 = 0 : \qquad g_0 &=& {1\over 2} ( 3 f_1^2 + 6 f_1 f_2 f_3 + f_2^3 -
2g_1 g_5 - 2g_2
g_4 - g_3^2) \nonumber \\
&=& {3\over 8} f_1^2 - {3\over
8}f_1 f_2f_3 -{1\over 16} f_2^3 + {9\over 64} f_2^2 
f_3^2  - {15\over 256} f_2 f_3^4 \nonumber\\
&& \qquad + {3\over
32} f_1f_3^3 + {7\over 1024} f_3^6 \equiv \wh g_0\, .
\een
This determines the parameters $g_i$ in terms of three independent
parameters $f_1,f_2,f_3$.  
In order to have an $\wh \E_2$ configuration we must also require $d_5$ to
be non-zero. A straightforward 
computation gives:
\be \label{ex25}
d_5 = {3\over 1024} (8 f_1 - 4 f_2 f_3 + f_3^3)(16 f_2^2 + 16
f_1 f_3 - 16 f_2
f_3^2 + 3 f_3^4) \, . 
\ee
Thus the most general properly isolated $\wh \E_2$ configuration is
labelled by three parameters $f_1,\ldots f_3$ satisfying the inequality
$d_5\ne 0$. Of these three parameters one is redundant due to the
freedom
of
scaling of $z$. Using this freedom, we can require the 
$d_5$ given in
\refb{ex25} to be equal to $(3/1024)$. This gives a two parameter family
of
$\wh \E_2$ configurations. In summary  
\begin{eqnarray}
\label{enhata}
 \widehat \E_2 &:&  \overline f (z)  = f_1 z + f_2 z^2 + f_3 z^3 +
z^4\,, 
\nonumber \\
&& \overline g (z) = \sum_{k=0}^6 \hat g_k z^k\, , \nonumber \\
&& (8 f_1 - 4 f_2 f_3 + f_3^3) (16 f_2^2 + 16
f_1 f_3 - 16 f_2
f_3^2 + 3 f_3^4) = 1\, .
\end{eqnarray}

The construction 
of properly isolated $\wh \E_n$ configurations with
$2<n<9$  
follows trivially. In this case we require $d_k$ to vanish for $k\ge
(n+4)$, and $d_{n+3}$ to be non-zero. Thus we need to satisfy
the first
$(9-n)$ of the equations \refb{ex18}-\refb{ex24}, and also require that
the left
hand side minus the right hand side of the $(10-n)$th equation
($g_{n-3}-\wh g_{n-3}$) be 
non-zero, which we can set equal
to $(+1)$
by using the freedom of rescaling $z$. Thus
the general solution is parametrized by $(n+1)$ parameters $f_1,\ldots
f_3,g_0,\ldots g_{n-3}$ subject to one `gauge fixing condition', which
determines
$g_{n-3}$ in terms of the other parameters. Using the gauge fixing 
condition $g_{n-3}-\hat g_{n-3}=1$,
the explicit
solution
is given by 
\begin{eqnarray}
\label{enhat}
 \widehat \E_{9>n> 2} 
: \quad \overline f (z)  &=& f_1 z + f_2 z^2 +
f_3
z^3 + z^4\,,
\nonumber \\
\quad \overline g (z) &=& \hat g_6\, z^6 + \cdots +\hat
g_{n-2} z^{n-2} + (1+ \hat g_{n-3}) z^{n-3} 
+ \sum_{k=0}^{n-4} g_k z^k \,,
\end{eqnarray}
where  $\hat g_6 = 1$, and the other $\hat g_n$ are given
in \refb{ex19}-\refb{ex24}. 
$g_0,\ldots g_{n-4}$ are arbitrary.

Let us now turn to the case of $\wh \E_1$. In this case we need to
satisfy
eqs.\refb{ex18}-\refb{ex24}, together with
$d_5 = 0$. It follows from  (\ref{ex25}) that we need
\be \label{ex26}
(8 f_1 - 4 f_2 f_3 + f_3^3)(16 f_2^2 + 16
f_1 f_3 - 16 f_2
f_3^2 + 3 f_3^4) =0\, .
\ee
Note that this equation contains two factors. Furthermore, it is
straightforward
to
see that if we require both factors to vanish simultaneously, then
$\Delta$ vanishes identically, and hence we have an unphysical solution.

Thus it appears that in the parameter space labelled by $f_1$, $f_2$,
and
$f_3$, there are two physically
disconnected regions which give 
properly isolated 7-brane configurations
with the same number of 7-branes and the same monodromy 
as the $\wh \E_1$
configuration:
\be \label{ex27}
(8 f_1 - 4 f_2 f_3 + f_3^3) = 0\, ,
\ee
or
\be \label{ex28}
(16 f_2^2 + 16
f_1 f_3 - 16 f_2
f_3^2 + 3 f_3^4) =0\, .
\ee
It turns out that among the configurations of 7-branes
listed in table 5 of \cite{9812209} there is precisely one more
configuration with the same monodromy and the 
same number of 7-branes as
the $\wh \E_1$ configuration, namely the $\wh{\wt \E}_1$ configuration.
Thus
we expect 
to identify one of the branches of \refb{ex26} 
with $\wh{\wt \E}_1$, 
and the other branch with  $\wh \E_1$.
Let us begin with the first branch, given by equation (\ref{ex27}) which 
we can
use to solve for
$f_1$ as
\be\label{solvef1}
f_1 = -{1\over 8} f_3^3 + {1\over 2} f_2 f_3 \,.
\ee
With this value of $f_1$ we can now find
\be \label{ex30}
d_4 =  - {3 \over 16384} (4 f_2 - f_3^2)^4\, . 
\ee
We have two parameters $f_2$ and $f_3$. It is convenient at this stage
to introduce a new parameter $s$ through the relation 
$f_2 = {1\over 4} s f_3^2$,
and then set $f_3 = 4$ as a gauge condition.\footnote{Note that 
this gauge condition is valid for all $(f_2,f_3)$ as long as $f_3\ne 0$,
{\it i.e.} $s\ne \infty$. If $f_3$ vanishes, then we need to choose a
different gauge.} We then have $d_4 = -12 (s-1)^4$, and therefore we
should get the desired configuration when $s\not= 1$.
At this stage we can write every coefficient in terms of
$s$.  
As we will explain shortly, this is $\widehat \E_1$. We then have 
\begin{eqnarray}  
\label{eonehat}
 \widehat \E_1 : \quad \overline f (z)  &=& z^4 + 4 z^3 + 4s\,  z^2
+ 8 (s-1)\, z  \,, \qquad s\not=1  
\nonumber \\
\quad \overline g (z) &=&  z^6 + 6 z^5 + 6 (s+1) z^4 + (24s-16)\, z^3 \\
&& \,\, + 6(s+3)(s-1)  z^2 + 12 (s-1)^2\, z - 4(s-1)^3\, . \nonumber 
\end{eqnarray}
{}From this one finds
\be\label{deltahate1}
\Delta (\widehat \E_1) = 4 (s-1)^4 \Bigl( -3z^4 - 12 z^3 - 12 s\,
z^2 - 24 (s-1)\,\,z + 4 (s-1)^2 \Bigr)
\ee
Let us now confirm that this is $\widehat \E_1$. 
To this end we  recall that the $\widehat \E_1$ brane configuration
is ${\bf B} {\bf C} {\bf B} {\bf C}$ and either the ${\bf B}$ or
the ${\bf C}$ branes can be brought together to define an $\A_1$
singularity.
Indeed,  we found that letting
\be\label{findsing}
z = -1 + y\,,\qquad  s = -{1\over 2} - \sqrt{3}
\ee
equations (\ref{eonehat}) and (\ref{deltahate1}) become
\begin{eqnarray}  
\label{eonehata}
 \overline f (y)  &=& (7+ 4\sqrt{3} )  - (8+ 4\sqrt{3}) y^2  + y^4
\nonumber \\
 \overline g (y) &=&  (26 + 15 \sqrt{3})  + (30\sqrt{3} + {105\over 2})
\, y^2 -(12 + 6\sqrt{3})\,y^4 +
y^6
\\
\Delta(y) &=&  -{27\over 4} (97 + 56\sqrt{3}) \, y^2 \, (y^2 - 4\sqrt{3}
-8)
\,.
\nonumber
\end{eqnarray}
This is an $\A_1$ singularity  at $y=0$; indeed, at this point we have
$\hbox{ord}(f) =
\hbox{ord} (g) =0$, and $\hbox{ord} (\Delta) = 2$.

This confirms that we are dealing with $\wh{\E}_1$. 
We can perform another check. It should not
be possible to decouple a brane in this configuration, since removing any
single brane from $\wh \E_1$ will leave a configuration with hyperbolic
monodromy \cite{9812209} and such configuration (by our earlier arguments)
cannot be isolated.
Indeed, to make the coefficient of $z^4$ in $\Delta$ vanish, we must take
$s=1$, but this makes $\Delta$ vanish identically, and therefore this is
not a physical
solution.\footnote{We also need to make sure 
that the special point $s=\infty$, where our gauge choice breaks down,
does not correspond to a decoupled brane configuration. To see this we go
back to eq.\refb{ex30} and set $f_3=0$. Requiring $d_4$ to vanish will now
require $f_2$ to vanish. This, in turn, makes $f_1$ and all the
coefficients $g_k$ for $0\le k\le 5$ to vanish. Thus we get $\wb f=z^4$,
and $\wb g=z^6$. This makes $\Delta=\wb g^2 - \wb f^3$ vanish
identically.}

We now begin the exploration of the second branch, indicated in
(\ref{ex28}). In here we
must set:
\be\label{ex100}
f_1 = - {f_2^2\over f_3} + f_2 f_3 - {3\over 16} f_3^3\,.
\ee
With this condition, we now examine the resulting value of $d_4$ which
turns out to be
\be \label{ex31}
d_4 = {3\over 16384} { (8 f_2 - 3 f_3^2) (4 f_2 - f_3^2)^4 \over f_3^2}
\, .
\ee
We have two parameters $f_2$ and $f_3$. It is convenient at this stage
to relate them via another parameter. 
We put $f_2 = {1\over 4} s f_3^2$,
and then to set $f_3 = 4$ as a gauge condition.\footnote{Again, 
this
gauge
condition breaks down if $f_3=0$.} We then find
\begin{eqnarray}  
\label{eonehattil}
 \widehat{\wt \E}_1 : \quad \overline f (z)  &=& z^4
+ 4 z^3 + 4s z^2 - 4 (s-1) (s-3) z\,,   \quad s\not= 1 
\nonumber \\
\quad \overline g (z) &=&  z^6 + 6 z^5 + 6(1+s) z^4 + ( -22 + 36s -6s^2
) z^3 \nonumber \\  
&& \,\, -6(s-1)(s-5)  z^2 -12(s-2)(s-1)^2\, z  \nonumber\\
&& \,\,+ 2(3s-5) (s-1)^3 \,.
\end{eqnarray}
and one can confirm that
\begin{eqnarray}
\label{de100}
\Delta (\widehat {\wt \E}_1) &=& (s-1)^4 \,\Bigl(\, 12 (2s-3)\,z^4 - 8
(s^2 -14s + 19)\,
z^3
\nonumber\\
&& \qquad\qquad + 24 (3s^2 - 4s -1)\, z^2 - 48 (3s-5)(s-2)(s-1) \, z \\
&& \qquad\qquad + 4 (3s-5)^2 (s-1)^2 \Bigr) \nonumber
\end{eqnarray}
This is the one-parameter presentation of $\wh{\wt \E}_1$.
To confirm this end we recall that  $\wh{\wt \E}_1$ is described as
${\bf A}{\bf X}_{[2,-1]} {\bf C} {\bf X}_{[4,1]}$, 
and the ${\bf A}$ and
${\bf C}$ branes, for example, can be 
brought together  at
$z=0$. For this one must have 
$f\sim z$ and $g\sim z$. Since $\overline f\sim z$ in the 
above, we must
see if it is possible to set to zero the $z$-independent term in
$\overline g$.
Indeed, we see two possibilities. The first  one, $s=1$ is ruled out,
since
then $\Delta$ vanishes identically. 
On the other hand we can take $s=
5/3$. This
gives
\begin{eqnarray}  
\label{eo200l}
 \overline f (z)  &=& z^4 + 4 z^3 +
{20\over 3}  z^2  +{32\over 9} z \,, 
\nonumber \\
\quad \overline g (z) &=&  z^6 + 6 z^5 + 16 z^4 + {64\over 3} z^3
+ {40\over 3}  z^2
+ {16\over 9}  z  \\
\Delta  &=& {64\over 81} \, z^2 \Bigl( z^2 + {28\over 9} z + 4 \Bigr)
\end{eqnarray}
This is indeed $\wh{\wt \E}_1$ with an $\H_0$ singularity at $z=0$.

Finally, we can identify the configuration $\wh{\wt \E}_0$ which has
three 7-branes, by decoupling a brane from $\wh{\wt \E}_1$. 
It is clear that we
must set to zero the coefficient of $z^4$ in $\Delta$ as given in
(\ref{de100}).
Taking $s=1$ is clearly illegal, so we must take $s= 3/2$. In this case
we find
\begin{eqnarray}  
\label{eot20}
 \overline f (z)  &=&  z^4 + 4 z^3 +
 6 z^2  +3 z \,, \\
\overline g (z) &=&  z^6 + 6 z^5 + 15 z^4 + {37\over 2} z^3
+ {21\over 2}  z^2
+ {3\over 2}  z  - {1\over 8} \,.
\end{eqnarray}
This can be simplified by letting $z \to z-1$. One then 
obtains:  
\begin{eqnarray}
\label{eot2ss}
\wh{\wt \E}_0 : \qquad \overline f (z)  &=&  z^4 - z \,,
\nonumber \\
\quad \overline g (z) &=&  z^6 - {3\over 2} z^3
+ {3\over 8} \\
\Delta  &=& -{1\over 8} z^3 + {9\over 64}\,.
\end{eqnarray}
The same shift $z\to z-1$ would also simplify somewhat
the presentations of $\wh{\E}_1$ and $\wh{\wt \E}_1$
given earlier.
This concludes our proof that all the $\wh
\E_n$ ($1\le n\le 9$) and $\wh{\wt \E}_n$ ($n=0,1$) configurations 
can be properly isolated.

\subsubsection{The orthogonal series $\D_n$ ($0\le n\le 4$)} 
\label{sx33}

The $\D_n$
configuration has monodromy $-T^{4-n}$
and has $(n+2)$ 7-branes.
The
appropriate $f$ and $g$ in these cases coincide with the corresponding
functions found in ref.\cite{9408099} for $\NN=2$ supersymmetric SU(2)
gauge theories with $n$ hypermultiplets in the fundamental
representation \cite{9605150}. However for completeness we shall
construct
these functions explicitly here, as it does not require any extra
effort.

Proceeding in the same way as in the $\wh \E_n$ case, we conclude that
the
$\D_n$ configuration for $n\le 4$ is described by 
polynomials $f$ and $g$
of degree 2 and 3 respectively, subject to the condition that $\Delta$
is
a polynomial of degree $(n+2)$. We introduce the coefficients $f_k$,
$g_k$ and $d_k$
through the equations:
\ben \label{ex35} 
\wb f(z) & \equiv &
 - (4)^{1/3} f(z) = \sum_{k=0}^2 f_k z^k \nonumber \\
\wb g(z) & \equiv & (27)^{1/2} g(z) = \sum_{k=0}^3 g_k z^k \nonumber \\
\Delta & \equiv & -\wb f^3+\wb g^2 =\sum_{k=0}^{6} d_k z^k\, .
\een
Then for $\D_4$ the only requirement on 
the coefficients is that $g_3^2\ne f_2^3$. In other words:
\be \label{ex150}
\de(f)\le 2, \qquad \de(g) \le 3\,,  \qquad \de(\Delta) = 6.
\ee
In order to get a
$\D_n$ configuration for $n\le 3$, $d_k$ must vanish
for
$k\ge (n+3)$, and $d_{n+2}$ should be non-zero. This gives constraints
on
the coefficients $f_n$ and $g_n$. As in the $\wh \E_n$ case,
by using the freedom of shifting $z$ and
rescaling $f$ and $g$
we set
\be \label{eex1}
f_0=0\, , \qquad   f_2=1\, .
\ee
Instead of discussing the case of
each of the $\D_n$'s separately,
it is most convenient to start with
$\D_0$. By equating the coefficients of $d_6, d_5, d_4 $ and $d_3$ to
zero we find the following  constraints:  
\be \label{ex36}
d_6 = 0 : \quad g_3 = 1 \equiv \wh g_3 \, ,
\ee
\be \label{ex37}
d_5 = 0 : \quad g_2 = {3\over 2} f_1 \equiv \wh g_2\, ,
\ee
\be \label{ex38}
d_4 = 0 : \quad g_1 = {1\over 2} (3 f_1^2 - g_2^2) = {3\over 8} f_1^2
\equiv \wh g_1 \, ,
\ee
\be \label{ex39}
d_3 = 0 : \quad g_0 = {1\over 2} (f_1^3 - 2 g_1 g_2) = -{1\over 16} f_1^3
= \wh g_0 \, .
\ee
These equations determine the coefficients $g_k$ in terms of the single
parameter $f_1$. Using eqs.\refb{eex1}-\refb{ex39} we get
$d_2 = -{3\over 64} f_1^4\,.$ 
This term must not vanish, and therefore 
$f_1\neq 0$. We can use $z$-scaling together with compensating
$f$ and $g$ scalings (as before) to fix $f_1 = 4$, while preserving
$f_4 = 1$.  We then have for $\D_0$ 
\begin{eqnarray}
\label{d0pre}
 \D_0 &:&  \overline f (z)  = z^2 + 4z \,, 
\nonumber \\
&& \overline g (z) = z^3 + 6z^2 + 6z -4\, , \nonumber \\
&& \Delta (z) = -12 z^2 - 48 z + 16 .
\end{eqnarray}

The solution for $\D_n$ for all
other $n\le 3$ is now easily obtained. 
For this the coefficients $g_n$  
need to satisfy
the first $(4-n)$ equations in eq.\refb{ex36}-\refb{ex39}, and should
not
satisfy the $(5-n)$th of these equations. 
This determines the parameters
$g_n, g_{n+1},\ldots g_3$
in terms of $f_1$, and gives a strict
inequality for $g_{n-1}$. As in the $\wh \E_n$ case, we can use the
freedom
of scaling $z$ to ensure that the difference between the left and the
right
hand side of the $(5-n)$th equation is 1. This gauge fixing
condition determines $g_{n-1}$. Thus
the general solution is parametrized by $n$ parameters $f_1,g_0,\ldots
g_{n-2}$. Explicitly, the answer is:
\begin{eqnarray}
\label{dnser}
 \widehat {\D}_{n> 0} : \quad \overline f (z)  &=& z^2 + f_1 z  \,,
\nonumber \\
\quad \overline g (z) &=& \hat g_3\, z^3 + \cdots +\hat
g_{n} z^n + (1+ \hat g_{n-1}) z^{n-1} + \sum_{k=0}^{n-2} g_k z^k
\end{eqnarray}
where  $\hat g_3 = 1$, the other $\hat g_n$ are given
in \refb{ex37}-\refb{ex39}, and 
$g_0,\ldots g_{n-2}$ are arbitrary.

This finishes 
explicit construction of all the properly isolated 7-brane
configurations. 

\subsection{Examples of asymptotically isolated 7-brane 
configurations} \label{sx34}

We now turn to the asymptotically isolated 7-brane
configurations.
As stated earlier, we shall not attempt to completely
classify or to give explicit constructions of all such
7-brane configurations. But we shall consider two examples.

$\D_n$ ($n> 4$): In this case
the monodromy, $-T^{4-n}$,  is proportional to
a negative power of $T$. From our analysis in the last 
section we see that the only way such a monodromy can be
obtained
is in case 4, where the overall coefficients in the functions $f$ and
$g$
blow up in the $\lambda\to 0$ limit. 
Thus these configurations cannot be
properly isolated. 
Indeed, 
for these
configurations the string coupling $(Im(\tau))^{-1}$ grows at large
distance, and in order to prevent it from blowing up at a finite value
of
$z$ (which would represent other 7-branes), the string
coupling
at finite points in the $z$ plane must be made smaller and smaller as
$\lambda$ approaches zero.
We shall not attempt to
give an explicit
construction of these configurations
here.
However, the existence of such
configurations
can be proved as follows. We
start with a collapsed
$\D_n$ configuration at $z=0$ which is known to exist for $n\ge 4$, and
resolve the
singularity slightly so that the branes in the $\D_n$ configuration are
located at $z\sim\eps$ for some small
number $\eps$. Now we can rescale $z$ by $(1/\eps)$ to put these branes
at
finite values of $z$. This takes all the other branes to large
values of $z$.

In order to verify that these configurations satisfy condition
\refb{exx4}, we note that at a $\D_n$ singularity $f(z)$ has a double
zero. Thus, after 
resolving the singularity
and rescaling, $f(z)$ has two zeroes at finite $z$. This gives 
$d_f=2$, 
which satisfies the inequality \refb{exx4}.

This proves the existence of asymptotically isolated $\D_n$
configurations
for $n>4$.

$\A_n$: This configuration has $(n+1)$ 7-branes and has monodromy
$T^{-(n+1)}$.  
Thus it also belongs to the class of 7-branes which can only 
be asymptotically isolated.
Again we shall not discuss explicit construction of these
configurations. The existence of such configurations can be proved
in the same way as the $\D_n$ case for $n>4$ by resolving a 
configuration with $\A_n$ singularity, followed by a scaling of $z$. For
an $\A_n$ singularity $f(z)$
has
no zeroes at the location of the singularity. Thus after resolution of
the
singularity and appropriate rescaling, there will be no zero of $f$ at
a finite value of $z$. This gives 
$d_f=0$, which satisfies \refb{exx4}. 

\sectiono{Brane Configurations with
Stable non-BPS states}
\label{sx4}

In the present section we investigate which 7-brane configurations
support \nbps. Such states take the form of string junctions
extending between the seven 
branes.\footnote{When we refer to a 
junction 
corresponding to a specific
vector in the junction lattice, it corresponds to the minimal mass
configuration among a whole set of junctions
which can be continuously deformed to each other by manipulations of the
form discussed in \cite{9709013} and \cite{9801205}.}
We will focus
on brane configurations that can be isolated
in the sense discussed in the
previous  sections.  Unless the brane configuration can be isolated
we cannot
reliably ascertain  the stability of the candidate states.
In a later
section we will discuss some aspects of non-isolable configurations
and their potentially \nbps.

Once we focus on a particular brane configuration, we only
examine string junctions joining 7-branes of that configuration.
We call these {\it localized junctions}, since they do not carry
away charge to some remote 3-brane or to another set of 7-branes.
We now claim that a string junction $\J$ on such 7-brane configuration
corresponds to a {\it possibly
stable non-BPS state} if:
\medskip

\noindent
(i) The associated homology cycle $\J$ satisfies
$\J^2 < -2$.

\medskip
\noindent
(ii) $\J \not= \sum_i n_i\j_i$, where $\j_i$
are homology cycles satisfying $\j_i^2 \geq -2$ and $n_i$ are
arbitrary integers.

\medskip
\noindent
Let us first examine the first condition. Recall that 
in F-theory on an elliptically fibered K3 over base $S^2$, a string
junction joining type IIB seven-branes on $S^2$ can be associated
to a two cycle in K3. This cycle,  being boundaryless,
corresponds
to an element of the second homology class of K3.
We use the symbol $\J$ to denote interchangeably  the junction and
the associated homology cycle. We also denote by $\J^2$ the
self-intersection number of the cycle. It is well-known that in K3
any cycle with $\J^2 = 2g-2$ ($g\geq 0)$ has a holomorphic
representative of
genus $g$. That representative defines a BPS junction. On the other hand,
when $\J^2 < -2$ the cycle has no holomorphic representative, and the
associated junction is never BPS. Thus the first condition guarantees
that the state is not BPS.\footnote{This can 
also be seen in the dual heterotic string theory on $T^2$ as follows. In
the heterotic description $\J$ corresponds to a vector on the Narain
lattice\cite{NARAIN} and $\J^2$ corresponds to its squared norm. Since in
the heterotic
string theory there are no BPS states with $\J^2<-2$, we see string
junctions with $\J^2<-2$ cannot be BPS.}

Let us now consider the second condition. Suppose $\J= \sum_i n_i\j_i$,
where
$\j_i$ are homology cycles satisfying $\j_i^2 \geq -2$ and $n_i$
are some integers. The equality
of homology cycles implies that whatever charges $\J$ carries they
are also carried by the total set of states associated to the
right hand side. Since all states in the right hand side are
BPS ( $\j_i^2\geq -2$) the decay of $\J$ into stable BPS states cannot
be ruled out by charge conservation. Therefore, condition (ii) ensures
that the state cannot decay into stable BPS states. Of course, 
even if condition (ii) is not satisfied, 
the non-BPS state will be stable
if its mass is lower than the sum of the masses of
the possible product states. But this requires a detailed study of the
masses of various states. We do not attempt to carry out such analysis
here.

In general, a brane configuration will admit many or infinite
number of possibly stable non-BPS states, namely states satisfying
conditions (i) and (ii). Such states may decay into each other,
but there will be at least one state $-$ the lightest of the possibly
non-BPS stable states $-$ or perhaps more that
will be  {\it genuinely stable non-BPS state(s)}.
Which particular states are
stable, and the number of such states, may vary as we change the
parameters labelling the isolated 7-brane configuration.

\medskip
Given a 7-brane configuration with $(N+2)$ branes, the
fact that the junction does not carry away charge imposes
two conditions (unless all branes are mutually local) $-$ one
corresponding to the D-string charge and another corresponding to the
fundamental string charge $-$ and therefore the set
of localized junctions  is spanned by $N$ linearly
independent junctions. If one identifies a semisimple
algebra of rank $N$ on this brane configuration there
cannot be \nbps. Indeed, having identified a rank $N$
semisimple algebra means
having identified a  set of $N$ linearly independent
localized BPS junctions representing the simple
roots of the algebra\cite{9812028,9812209}.
This is therefore a basis for
the set of all
localized  junctions, and therefore any junction can be
written as some integral linear combination of junctions
that are BPS, in violation of  condition (ii).  What
one needs is $u(1)$ factors in the algebra carried by the 
branes. 
Such factors arise when holomorphic junctions do not span the lattice of
localized junctions. 

We shall first examine the basic realizations of the
(extended) $\A$, $\D$, $\E$, $\widehat \E$ and $\H$ 
series and find the cases
that can give rise to  possibly stable non-BPS states.
We follow the convention of refs.\cite{9812028,9812209} of denoting
by $\X_{p,q}$ the $(p,q)$ 7-brane with monodromy $K=\pmatrix{1+pq & -p^2\cr
-q^2 & 1-pq}$, and define special 7-branes $\A$, $\B$ and $\C$ as
$\A=\X_{[1,0]}$, $\B=\X_{[1,-1]}$ and $\C=\X_{[1,1]}$. The monodromy of a
brane
configuration containing a product of $\X_{p,q}$'s is obtained by
multiplying the individual monodromy matrices in opposite order.

\medskip
\noindent
$\bullet$  The $\A_N$ series ($N\geq 1$).
These configurations are special in that all branes are mutually local;
the configuration
is produced by $(N+1)$ $\A$ branes.
With just one charge conservation
condition corresponding to the fundamental string charge, localized
junctions are spanned by $N$ basis elements.
The $N$ junctions joining $\A_i$ to
$\A_{i+1}$ represent the $su(n+1)$ roots and span the
lattice of localized
junctions. Thus (ii) cannot be satisfied.

\medskip
\noindent
$\bullet$  The $\D_N$ series ($N\geq 0$).
These brane configurations are
${\bf D}_N =\A^N
\B\C$. For the case $\D_0$, which only has two branes, there
are no localized junctions. The configuration
$\D_1$ carries a $u(1)$ algebra only and thus
has a candidate non-BPS state. On the other hand for
$\D_{N\geq 2}$ the algebra is semisimple and therefore there are no
possibly stable non-BPS
states.

\medskip
\noindent
$\bullet$  The $\H_N$ ($N\geq 0$). Only $\H_{N\leq 3}$ can be isolated.
($\H_3 = \D_3$). This series is  realized as
${\bf H}_N =
\A^{N+1}\C$ and the algebra is semisimple for all $N\geq 1$. The
remaining case,
$\H_0$, has no localized junctions.

\medskip
\noindent
$\bullet$  The $\E_N$ series.  Here we have the two realizations
$\E_N = \A^{N-1}\B\C\C$ or $\tilde\E_N = \A^N \X_{[2,-1]}\C$, which
are equivalent for $N \geq 2$. Here  $\E_5 (= \D_5)$,
$\E_6, \E_7,\E_8$ and
$\E_9
= \widehat \E_8$ can be isolated. On the other hand all these
give semisimple algebras so (ii) is not satisfied.

\medskip
\noindent
$\bullet$  The $\widehat \E_N$ series.  Once more
we have the two
realizations
$\widehat\E_N = \A^{N-1}\B\C\C\X_{[3,1]}$ ($N\geq 1$) and
$\widehat{\tilde\E}_N =
\A^N
\X_{[2,-1]}\C\X_{[4,1]}$ ($N\geq 0$),
which are again equivalent for $N \geq
2$. All these configurations can be isolated at least 
for $N\le 9$, 
and correspond to
parabolic monodromies. For
$N\geq 3$ one identifies semisimple affine algebras and thus no
possibly stable non-BPS
states. Both $\widehat \E_2$ and $\widehat{\tilde \E}_1$ carry
affine $u(1)$ factors and thus are candidates for having possibly \nbps.

\bigskip
\noindent
In conclusion, the conditions of isolation, plus
(i) and (ii) have restricted the list to  the cases of
$\D_1,  \widehat{\tilde \E}_1, \hbox{and} \,\widehat \E_2$.
We will now examine these cases in detail and confirm that they
have collections of possibly \nbps, and therefore some
genuinely stable non-BPS states.

\subsection{Case of $\D_1$.}~ 
The configuration here is $\A\B\C$
and having  mutually nonlocal branes the lattice of localized junctions
is one dimensional. This case is identical to a D7-brane O7-plane
system analyzed in ref.\cite{9803194}. Using the conditions of charge
conservation
one readily finds that this lattice is spanned by the minimal proper
junction $\J = 2 \a - \b -\c$. Here we are following
the convention of \cite{9812028,9812209} that a junction ${\bf
x}_{[p,q]}$ denotes a $(p,q)$ string departing from the ${\bf X}_{[p,q]}$
7-brane and going to $\infty$. 
We can easily verify that
$\J$ satisfies conditions (i)
and (ii). Indeed, using the rule\cite{9804210} that each elementary
junction ${\bf x}_{[p,q]}$ has self-intersection $-1$, and that the
intersection number of a junction ${\bf x}_{[p,q]}$ with another
junction $\x_{[p',q']}$ to its right is ${1\over 2}(pq'-qp')$, we
get
$\J^2 = -4$, and therefore condition (i) is satisfied. (The junction $\J$
corresponds to two  strings departing the $\A$ brane
and
meeting
after  going around the $\B$ and $\C$ branes.
In this picture the self-intersection is manifestly $(-4)$.)
Since any localized junction must be a multiple of the minimal junction
$\J$,
any junction must satisfy condition (i) and therefore there are no
BPS junctions
in this configuration. As a consequence condition (ii) is
also satisfied.  The states $(n\J)$, for $n\not=0$ are all possibly stable
non-BPS states. 
In the limit when the separations between the 7-branes are large compared
to the string length scale, the mass of a string junction can be computed
reliably by
integrating the tension along the various segments of the junction.
In this classical limit the minimal mass configuration 
in the class of  $n\J$
corresponds to $n$ copies of the minimal mass junction in 
the class of  $\J$. Thus
the mass of the former is approximately $n$ times the mass of the latter.
Thus the minimal charged states $\pm
\J$ are genuinely stable non-BPS. These were identified in \cite{9803194}.

\subsection{Case of $\widehat{\tilde\E}_1$} This is the brane
configuration having the following four seven branes (ref.\cite{9812209},
eqn.(3.10))
\be
\label{e1til}
\widehat{\tilde\E}_1 = \A\X_{[2,-1]} \C \X_{[4,1]} = \A
\widehat{\tilde\E}_0 = \tilde\E_1 \X_{[4,1]}
\ee
We have written it in two ways; as an enhancement of
$\widehat{\tilde \E}_0$, and as an affinization of $\wt \E_1$.
Given that we
have four seven-branes we must have two junctions spanning the lattice of
localized junctions. We claim
that the following is a basis for localized junctions of
$\widehat{\tilde\E}_1$:
\begin{eqnarray}
\label{thejun}
\bar\J &=& 3 \a - \x_{[2,-1]} -\c\,, \,\,\qquad \bar\J^2 = -8\\
\d &=& \x_{[2,-1]} + 2\c - \x_{[4,1]}\,,\qquad \d^2 = 0 \,, \quad
\d \cdot \bar\J = 0\,.
\end{eqnarray}
Linear independence is manifest, $\bar\J$ is supported on the $\A$ brane
while
$\d$ is not, $\d$ is supported on the $\X_{[4,1]}$ brane, 
while $\bar\J$ is
not. After imposing the constraint that no D- or fundamental string
charge flows to infinity, any arbitrary junction
$p\a+q\x_{[2,-1]}+r\c+s\x_{[4,1]}$, with integers
$p,q,r,s$, can be expressed as $-(r+2s)\bar\J -s\delta$.
This
establishes that $(\bar\J,\delta)$ form a basis for the localized
junctions
of
$\wh{\wt \E}_1$.

The physical interpretation of these junctions can be found
by considering the subconfigurations.  Indeed, the localized 
junctions of $\wt \E_1 = \A\X_{[2,-1]} \C$
make a one-dimensional lattice  spanned by $\bar\J$.
On the  other hand the junction $\d$ can be presented as 
string loop of charge $(p,q) = (-1,0)$ surrounding the
configuration $\widehat{\tilde {\bf E}}_0$ (ref.\cite{9812209}, eqn.(3.11)).
This picture makes it manifest that $\d^2=0$, a fact that guarantees that
this
junction arises from a holomorphic cycle of genus one 
and is
therefore  
BPS \cite{9809026}. Having charge $(-1,0)$ the junction $\d$ can be moved
across the remaining $\A$ brane and be presented
as a loop surrounding the complete $\widehat{\tilde \E}_1$
configuration.\footnote{In fact such loop is the loop of the
$\widehat{\tilde \E}_N$ configuration for all $N$.}
This makes $\bar\J\cdot \d=0$  manifest.

Since the arbitrary junction $\J_{Q,\ell} = Q \bar\J + \ell \d$ satisfies
$\J^2_{Q,\ell} = -8Q^2$, no junction with support on the $\A$ brane can be
BPS.  $\d$ is  the basis of BPS junctions.
Thus the junctions $\J_{Q,\ell}$ with $Q\not=0$ are all
possibly stable non-BPS states. 
Among all such states there will be at least one lowest mass
state $\J_{Q_0,\ell_0}$ that is a genuinely stable non-BPS state.
The precise
value of $(Q_0,\ell_0)$, however, is not determined by this argument. 
In fact,
as we change the parameters labelling the isolated $\wh{\wt \E}_1$
configuration,
the values of $(Q_0,\ell_0)$ can undergo discrete jumps.
Note that for  
a fixed $Q$ the non-BPS states $\J_{Q,\ell}$ for all values of $\ell$
generate
a (level zero) representation of the $\widehat{u(1)}$ algebra carried by
the $\wh{\wt \E}_1$ configuration. Since this configuration is
non-collapsible
the affine symmetry is only spectrum generating, and states with different
values of $\ell$ will typically have different masses and different 
stability
properties.

\subsection{Case of $\widehat\E_2$}
 This is the configuration
described in ref.\cite{9812209},
eqn.(3.7):
\be
\label{tildee2}
\widehat{\E}_2 = \A\B\C \C \X_{[3,1]} =  \E_2 \X_{[3,1]} = \A\B\C\B\C
\ee
expressed also as the enhancement of $\E_2$.
Given that we have 5 seven-branes
we expect three junctions to span the lattice of localized junctions.
We claim that the
basis of three
junctions can be chosen to be
\begin{eqnarray}
\label{hjg}
\j =&& \c_1 -\c_2\,, \\
\J_- =&& 2 \a - \b -\c_1\,, \\
\d' =&& \b + \c_1 + \c_2 - \x_{[3,1]} \,,
\end{eqnarray}
with
\be \label{exinter}
\j^2=-2, \quad \J_-^2=-4, \quad (\d')^2=0, \quad \j\cdot\J_-=1,
\quad\j\cdot \d'=\J_-\cdot\d'=0\, .
\ee
To see that 
this is a basis, 
we note that an arbitrary junction of the
form
$p\a+q\b+r\c_1+s\c_2+t\,\x_{[3,1]}$, with integers $p,q,r,s,t$, and
satisfying the
condition for D- and fundamental string charge localization, can be
expressed as $-(r+s+2t)\J_--(s+t)\j-t\d'$.

The set of 
localized junctions of the $\E_2$
sub-configuration 
is spanned by
$\J_-$ and $\j$. This carries an
$su(2)\times
u(1)$. $\j$ corresponds to 
the root of
the $su(2)$ factor and 
$\J_-$ is associated to the $u(1)$ factor.
We also introduce $\J_+ = 2 \a - \b -\c_2$.
The states $\J_\pm$ form a doublet of the $su(2)$ satisfying
\be
\label{doeubl}
\J_\pm \cdot \j = \mp 1\,, \quad \J_\pm^2 = -4
\ee
The $\d'$ junction is a $(-1,0)$ loop
surrounding the configuration (ref.\cite{9812209}, eqn.(3.9)). This
explains why
$\j \cdot \d' = \J_-\cdot\d' = 0$.

We now claim that  no
junction in $\wh\E_2$ with support in the $\A$ brane can be
BPS. Indeed, with
\be
\label{anly}
\J_{n,m,\ell} = n \J_- + m\j + \ell\d'\quad\to\quad  \J^2 = -4n^2 -2m^2 +
2mn=-(m-n)^2-3n^2-m^2\, . 
\ee
Thus $\J^2<-2$ for any non-zero integer $n$. 
Thus junctions in
$\wh\E_2$ with support on the $\A$ brane satisfy (i) and (ii), and
are possibly stable  non-BPS states. The $u(1)$
charge is measured by the number of prongs on the $\A$ brane.

The $su(2)$ symmetry is exact when 
the two $\C$ branes coincide. In this case $\J_\pm$ would make 
a doublet of possibly stable
non-BPS states.
The possibly non-BPS $su(2)$ singlet of minimal $u(1)$ charge
 is readily shown to be the junction $\J_0 = 2\J_-+\j+\ell\d' = 4\a -
2\b -\c_1-\c_2+\ell\d'$. It carries twice the $u(1)$ charge of
any member of the doublet. 
We can construct possibly non-BPS states in higher representations
of $SU(2)$ in
a similar manner. {}From the structure of the lattice it is easy to see that
odd
values of the $u(1)$ charge must be associated to $su(2)$
representations in the conjugacy of the doublet, while even $u(1)$
charges must be associated to $su(2)$ representations in the conjugacy
class of the adjoint.
Once we fix a $u(1)$ and
an
$su(2)$ representation, the states $\J_{n,m,\ell}$ for all values of
$\ell$ generate a (level zero) representation of the affine
$(\widehat A_1 \oplus \widehat {u(1)} / \sim) $ algebra of the
$\widehat \E_2$ configuration. Which of these configurations represent
genuinely stable non-BPS state is a detailed dynamical question which we
shall not address.

Finally we note that this configuration represents a single D7-brane near
a pair of O7-planes. This is seen in the last presentation given
in (\ref{tildee2}).
The $\A$ brane represents a D7-brane, while each of the $\B\C$ factors
represents an O7-plane.

\sectiono{Non-BPS states on Non-isolable  Configurations}
\label{sx5}

The strategy that we have used so far in our search for stable non-BPS
states consists of two steps. First we need a subset of 7-branes in
F-theory on K3 such that there are non-BPS junctions living on this subset
of branes which are stable against decay into other states living inside
the same subsystem. Second, we need to ensure that these states are also
stable against decay into junctions with one or more prongs on the
7-branes external to this subsystem.

The second condition requires that this subset of branes can be isolated
and was the subject of study in sections \ref{sx2} and \ref{sx3}. In this
section we shall search for 7-brane configurations which satisfy the first
condition and not the second. This would ensure that the non-BPS states
living on this subsystem are stable against decay into BPS states living
on the same subsystem, but could be unstable against decay into junctions
with prongs on the external 7-branes. At present the significance of such
brane configurations is not totally clear. However these configurations
could be the starting point in our search for 7-brane configurations which
admit non-BPS states which are stable due to dynamical reasons, namely
that their mass is smaller (but not much smaller) than the possible decay
products.

We shall begin by discussing two examples which we already encountered in
section \ref{sx4}. The first example will be that of an $\wt\E_1$
configuration. This is generated by $\bar\J$ defined in eq.\refb{thejun}.
Since $\bar\J^2=-8$, any state of the form $n\J$ is non-BPS. The minimum
mass state in this family will be stable against decay into other states
living solely inside the $\wt \E_1$ brane system. When the relative
separation between the branes is large, this corresponds to the junction
$\pm\bar\J$. From \cite{9812028} we know that for $\wt \E_1$,
$Tr(K)=-6$. Since the monodromy is hyperbolic, the $\wt\E_1$ configuration
cannot be isolated.

The second example is that of $\E_2$. This is generated by the junctions
$\J_-$ and $\j$ defined in eq.\refb{hjg}. Any junction of the form
$\J_{m,n}=n\J_-+m\j$ has $\J^2=-(m-n)^2-3n^2-m^2<-2$ for $n\ne 0$. Thus
there should be at least one non-BPS state on this system with component
along $\J_-$ which is stable against decay into other states living solely
inside $\E_2$. Again we see from ref.\cite{9812028} that in this case
$Tr(K)=-5$, and hence this configuration cannot be isolated.

We shall consider two more examples. The first example will be
an arbitrary configuration of three seven branes.
The second example will be that of a four 7-brane configuration carrying a
$u(1)\times u(1)$ algebra.

\noindent
\underbar{Non-BPS states on three 7-branes}~ By an SL(2,Z)
transformation, any three 7-brane configuration can be put
in  the form
$\A\X\X'$ where
$\X$ is a $[p,q]$ brane, and $\X'$ is a $[p',q']$ brane. We also require
$q\not= 0$ and $q'\not= 0$,
as well as $[p,q] \not= [p',q']$, for otherwise we have at least
two mutually local branes and there will be BPS states carrying this
U(1) charge into which a possible non-BPS state can decay.
Without loss of generality we can
also assume that
both $q$ and $q'$ are positive. Define
\be \label{eneweq}
\Delta = pq' - qp'\, . 
\ee
The general junction is
\be
 \J = Q_A \a + Q \x + Q' \x'\,,
\ee
and using charge conservation to solve for $Q_A$ and $Q'$ in terms of
$Q$, we find
\be
\J =  - {Q\Delta\over q'}\,\a + Q\, \x - {Qq\over q'} \, \x'\,. 
\ee
Since $Q\Delta/q'$ and $Qq/q'$ could be fractional, 
this junction is not necessarily proper. To address this issue, let us
define $\ell = \hbox{gcd} (q,  q')$  and let $q'= \ell
q_0$. One must then choose $Q=q_0$ to get the minimal proper junction.
Indeed, this gives $Q/q' = 1/\ell$ and we get
\be
\J = {1\over \ell} \Bigl( -\Delta\,\a + q' \x - q \, \x'\Bigr) \,. 
\ee
The self-intersection is readily found to be
\be
\J^2 = -{1\over \ell^2} \Bigl( \Delta^2+ q'^2 + q^2 + qq'\Delta\Bigr)
\,,\quad \ell= \hbox{gcd} (q,q'), \quad \Delta = pq'-qp'\,.
\ee
Each of the terms contributing to $\J^2$ is now an integer. We can easily
choose $p$, $q$, $p'$, $q'$ 
such that $\J^2$ given above is $<-2$, so
that all
charged states living on this brane system are non-BPS states. The
lightest of them will be stable against decay into other states within
this system.

It is useful to write this in a more symmetric form. Let
$\z_i$
denote as usual the elementary junction joining the $i$-th 7-brane to 
$\infty$. Using the notation of \cite{9812028}, sect.2.1,
we define
$\z_{ij} = \z_i \times
\z_j=(p_iq_j-q_i p_j)$, where $(p_i,q_i)$ denotes the $i$-th 7-brane.
Thus we have, in the present case,
$\z_{12}= q,\,\,
\z_{23}=
\Delta,\,\,
\z_{31}= -q'$,  and moreover $\ell = \hbox{gcd} (\z_{12}, \z_{23} ,
\z_{31})$.
We thus have
\be \label{enewereq}
\J^2 = -{1\over \ell^2} \Bigl( \z_{12}^2+ \z_{23}^2 + \z_{31}^2
- \z_{12} \z_{23} \z_{31}\Bigr) \, .
\ee
Eqn.~(2.7) of \cite{9812028}  for three 7-branes gives:
\be \label{enewest}
\hbox{Tr} K =  2  -\z_{12}^2- \z_{23}^2  - \z_{31}^2
+ \z_{12} \z_{23} \z_{31}\, .
\ee 
{}From this
we see that
\be \label{enewest1}
\J^2 = {1\over \ell^2} \Bigl(\hbox{Tr} K - 2\Bigr) \,.
\ee
Since in order to get a non-BPS
junction  we need
$\J^2 < -2$, we must have
\be \label{eagai}
\hbox{Tr} K < 2(1-\ell^2)\,.
\ee
Thus we see that except when $\ell = 1$, all three 7-brane
configurations with stable non-BPS states will have
$\hbox{Tr} K < -2$, and thus have negative hyperbolic monodromies.
When $\ell =1$, one must  have $\hbox{Tr} K < 0$.
As table 5 of \cite{9812209} indicates, there is no three 7-brane
configuration with $\hbox{Tr} K = -1$, $-$ a fact that is not hard to
prove.
The isolable configuration $\D_1$
corresponds to the case $\hbox{Tr} K = -2$.
The non-isolable three 7-brane
configuration
$\tilde \E_1$  corresponds to the case $\hbox{Tr} K = -6$.

\noindent
\underbar{A case with $u(1)\times u(1)$}~ While refs.
\cite{9812028,9812209} mostly searched for configurations
with large symmetry algebras, it is clearly possible to put
together many branes and still fail to find any enlarged
semi-simple algebra. 
In such cases we must get $u(1)$ factors. We illustrate this
by considering a brane configuration  with four 7-branes, which has no
charged BPS states. 
The brane configuration is
\be \label{efourbrane}
\X_{[1,2]} \, \A\B\C\, ,
\ee
which is obtained by adding the $\X_{[1,2]}$ brane to the $\D_1$
configuration.
Since the lattice of localized junctions should
be two-dimensional we expect to find two $u(1)$'s.
One  can show that the following is a basis for
localized
junctions:
\begin{eqnarray} \label{efin}
\J_1 &=& -2\a +\b+\c \\
\J_2 &=& -\a +\b -\c +\x \
\end{eqnarray}
These satisfy
\be \label{efinp}
\J_1^2 = -4 \,, \quad \J_2^2 = -4\,, \quad \J_1 \cdot
\J_2 = 0\, ,
\ee
and therefore there are no BPS states on this brane configuration carrying
either of these U(1) charges.
This shows that this brane configuration has non-BPS junctions which are
stable against decay within this brane system.
The general localized junction on this brane configuration would
be $\J = Q_1 \J_1 + Q_2 \J_2$, where $Q_1$ and $Q_2$ are the
two $u(1)$ charges of the junction. For this configuration
$\hbox{Tr} K = -14$, confirming that it cannot be isolated.

\bigskip
\noindent{\bf Acknowledgements.} A.S. would like to thank the
Center for Theoretical Physics at MIT for hospitality during
part of this work, and the participants
of the ICTP workshop on String Theory 
for providing a stimulating environment during the
course of this work. B.Z. would like to thank the Department
of Physics at Harvard University, for hospitality during the
concluding stage of this work.

The work of B.Z. is supported by the U.S. Department of Energy
under contract \# DE-FC02-94ER40818.

\end{document}